\documentclass[10pt, conference,letterpaper]{IEEEtran}

\IEEEoverridecommandlockouts
\usepackage{cite}
\usepackage{amsmath,amssymb,amsfonts}
\usepackage{algorithmic}

\usepackage{textcomp}
\usepackage{xcolor}
\def\BibTeX{{\rm B\kern-.05em{\sc i\kern-.025em b}\kern-.08em
    T\kern-.1667em\lower.7ex\hbox{E}\kern-.125emX}}

\usepackage{graphicx}
\usepackage{subfiles}
\usepackage[inline]{enumitem}

\usepackage{array} 
\usepackage{url}
\usepackage{epstopdf}
\usepackage{tikz}
\usetikzlibrary{automata,arrows,positioning,calc,%
shapes,chains}

\usepackage{mathtools}
\usepackage{texdef2015}
\allowdisplaybreaks

\newcommand{\Lcal}{\mathcal{L}}
\newcommand{\Qcal}{\mathcal{Q}}

\newcommand{\R}{\mathbb{R}}

\newcommand{\ql}{q_l}

\newcommand{\laml}[1][l]{\lambda^{(#1)}}

\newcommand{\smvec}[1]{\left[\begin{smallmatrix} #1\end{smallmatrix}\right]}
\newcommand{\rowvec}[1]{[\begin{matrix} #1\end{matrix}]}
\newcommand{\rvec}[1]{\smrvec{#1}}
\newcommand{\smrvec}[1]{\setlength\arraycolsep{2pt}\rowvec{#1}}

\newcommand{\piv}{\text{\boldmath{$\pi$}}}

\newcommand{\pibar}{\bar{\pi}}
\newcommand{\vvbar}{\bar{\vv}}
\newcommand{\vvhat}{\hat{\vv}}

\newcommand{\onev}[1][n]{{\mathbf{1}}_{#1}}

\renewcommand{\vec}[1]{\begin{bmatrix} #1\end{bmatrix}}

\newcommand{\vvmbar}[2][m]{\bar{\vv}_{#2}}
\newcommand{\vvmhat}[2][m]{\hat{\vv}_{#2}}

\newcommand{\age}{\Delta}

\newlength{\swwidth}

\newcommand{\lam}{\lambda}
\newcommand{\lamhat}{\hat{\lambda}}
\newcommand{\lamsum}{\lambda^*}

\newcommand{\sig}{\sigma}

\newcommand{\mh}{\hat{\mu}}
\newcommand{\musum}{\mu^*}
\newcommand{\bt}{\beta}
\newcommand{\xh}[1]{\hat{x}_{#1}}

\newcommand{\gp}{4}

\newcommand{\rh}{\hat{\rho}}

\newcommand{\locup}{location update }
\newcommand{\apup}{app update }

\newcommand{\RCU}{\text{RCU}}
\newcommand{\RWL}{\text{RWL}}

\begin{document}

\title{Lock-based or Lock-less: Which Is Fresh?}

\author{\IEEEauthorblockN{Vishakha Ramani,
Jiachen Chen,
Roy D. Yates}
\IEEEauthorblockA{WINLAB, Rutgers University \\
Email: \{vishakha, jiachen, ryates\}@winlab.rutgers.edu
}}





\maketitle
\begin{abstract}
We examine status updating systems in which time-stamped status updates are stored/written in shared-memory. Specifically, we compare Read-Copy-Update (RCU) and Readers-Writer lock (RWL) as shared-memory synchronization primitives on the update freshness. To demonstrate the tension between readers and writers accessing shared-memory, we consider a network scenario with a pair of coupled updating processes. Location updates of a mobile terminal are written to a shared-memory Forwarder Information Base (FIB) at a network forwarder. An application server sends ``app updates'' to the mobile terminal via the forwarder. Arriving app updates at forwarder are addressed (by reading the FIB) and forwarded to the mobile terminal. If a FIB read returns an outdated address, the misaddressed app update is lost in transit. 
We redesign these reader and writer processes using preemption mechanisms that improve the timeliness of updates. We present a Stochastic Hybrid System (SHS) framework to analyze location and app update age processes and show how these two age processes are coupled through synchronization primitives.  Our analysis shows that using a lock-based primitive (RWL) can serve fresher app updates to the mobile terminal at higher location update rates while lock-less (RCU) mechanism favors timely delivery of app updates at lower location update rates.
\end{abstract}






\section{Introduction}

In a broad range of cyber-physical systems, a source generates {\em status updates}, time-stamped measurements of a random process of interest,  that  are sent to one or more destinations through a network. In 
addition to requiring low latency transmission of each update, these applications require timeliness in the status update process at these destinations. For example, 
in 
robotic telesurgery, a surgeon needs timely feedback on the robotic arm since stale feedback 
can lead to organ injuries and other ramifications \cite{dtremotesurgery},\cite{surgendoscopy}.

It therefore becomes imperative to 
scrutinize current network architectures in terms of their feasibility to satisfy  stringent real-time and freshness requirements. 
Delays of tens of nanoseconds (e.g., a sequential SSD write of 8 kB takes 1ms, a sequential SSD read of 8kB takes 1$\mu$s \cite{napkinmath}) can significantly degrade system performance, particularly when they accumulate across multihop network paths. Hence, understanding the behaviour of a single network node is crucial in the development of ultra-low latency networks that deliver timely information. 

On the other hand, 
advances in high performance computing have led to realizations of solutions to a range of problems  in engineering, medical sciences, space research etc.~that typically involve performing computationally intensive tasks in a short amount of time. This is achieved by a mix of technologies, including shared memory multiprocessor systems (for parallel computing), and vector processing along with various algorithms built on such architectures. A plethora of applications benefit 
from parallelization of various operations,
including, for example, high throughput transaction processing in distributed databases \cite{DeWhitt} and faster training employing embarrassingly parallel
processes in machine learning \cite{Ben-Nun}. 

Such applications with high inter-processor communication demands expose synchronization between multiple processors as a key bottleneck in parallel computation in terms of scalability and computation times. In particular, a critical success factor in shared memory multiprocessors is synchronization, namely the coordination of concurrent tasks to ensure data consistency and correctness. This issue concerning readers-writer concurrency is manifested in various places. For instance, in a distributed database system, the challenge is to prevent database updates performed by one user from interfering with database retrievals and updates performed by another \cite{Bernstein}. In parallel machine learning, wherein there is an equal partitioning of data points across available processors, each having access to some \textit{global state} (for e.g. model parameters), then an incorrect modification of global state could potentially conflict with operations on other processors\cite{Pan}.

For shared memory systems,
primitives  that enforce synchronization among reader threads and writer threads when accessing a shared resource are a way to avoid race conditions caused by concurrent manipulation by different writer threads. Specifically, these primitives allow multiple threads (readers and writers) to execute concurrently and ensure that results of reading and writing are predictable.

The existing literature on synchronization techniques focuses mostly on the algorithm, implementation and throughput performance (operations per unit time) in the critical sections\footnote{Formally, a critical section is a protected section of the shared resource that is protected against multiple concurrent accesses} \cite{gramoli2015, davidEPFLasynchronized, clements2012scalable}. However, there has been a lack of study on the impact of synchronization primitives on the timeliness of the data accessed from shared memory in 
real-time IoT systems \cite{abd2roleAoI, kimcyber}.
By contrast, there have been many studies addressing the timeliness of status updates in queues and networks  using {\em Age of Information} (AoI) freshness metrics; see the surveys  \cite{kosta2017age,Yates-SBKMU-2021jsac-survey} and references therein. 

Age of information is an end-to-end performance metric for the timeliness  of a status updating process. An update with time-stamp $u$ is said to have {\em age} $t-u$ at a time $t \geq u$ and the freshness of an update is decreasing with its age. 
A monitor receiving a stream of updates has an age process $\age(t)=t-u(t)$ when $u(t)$ is the time-stamp of the freshest received update. The recent AoI literature has focused on high-level systems such as  sensor networks, wireless multiple access channels, and networks of queues in which delays are typically measured in milliseconds. 

In this study, we employ AoI freshness metrics to compare the performance of status updating systems employing Read-Copy-Update (RCU) and Readers-Writer lock (RWL) synchronization primitives to disseminate information using shared memory. 
While sub-microsecond delays may be typical in shared memory reads and writes, the frequency of such operations  contribute to latency in  network switches and routers. Furthermore, due to advances in both link and radio access network technologies, the transmission time of packets is often negligible. We contend that in many applications the bottleneck in status updating systems has shifted from data communication over links 
to data storage, processing and retrieval in network nodes. It is therefore imperative to analyse how network devices affect the freshness of updates. With this motivation,
we 
analyze
two fundamental synchronization primitives --- Read-Copy-Update (RCU) and Readers-Writer Lock (RWL) --- to aid both network engineers and software developers 
in designing architectures and software libraries for 
applications requiring timely status updates. 
\section{System Model}
In this work, we focus on a  class of system 
in which a source generates time-stamped {\em updates} (i.e., measurements of a random process of interest) and a concurrent/shared data structure stores these  updates\footnote{The concurrent data structures usually reside in shared memory that is an abstract storage environment.}. While the system will have many sources, our focus will be on the shared data structure that tracks  the status  of a single process of interest. Going forward, we refer to the shared data structure as {\em shared memory}, or simply as {\em memory}. 
Hence, a writer is responsible for recording  the fresh source/sensor measurements  as updates in the memory. 
A reader fulfills clients' requests for these  measurements by reading from the memory.

\subsection{Timely Update Forwarding}
\label{sec:networkforwarding}
\begin{figure}[t]
    \centering
\includegraphics[width=0.49\textwidth]{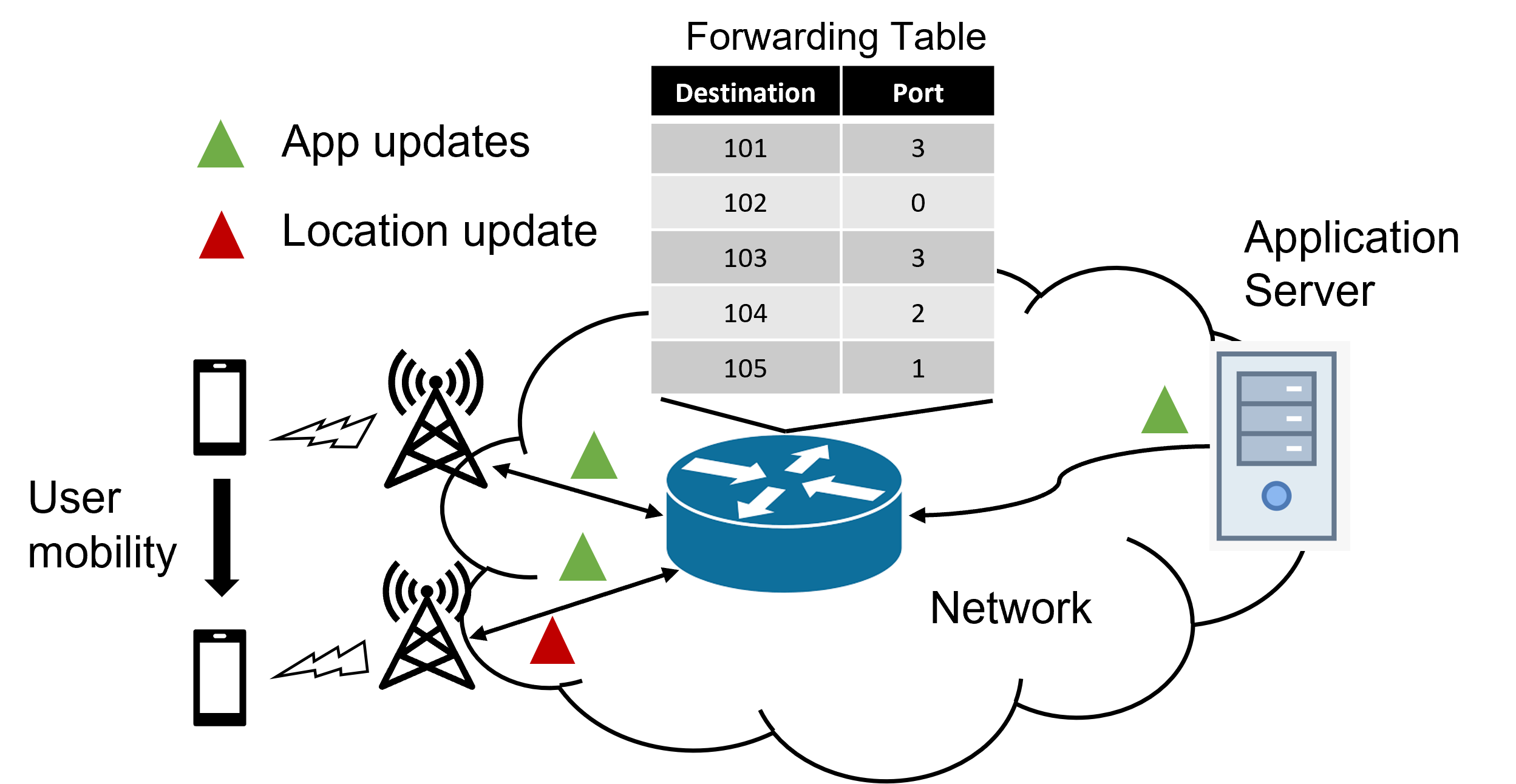}
    \caption{Packet forwarding application with mobile users}
    \label{fig:experimentconcept}
    \vspace{-5mm}
\end{figure}

To demonstrate the effect of concurrency constructs on 
timeliness,
we consider an example of packet forwarding in a mobile user environment, as shown in Fig.~\ref{fig:experimentconcept}. An application server in the network is sending ``app updates'' regarding a process of interest to a mobile terminal. The application sends its update packets to a forwarding node in the network. This forwarder maintains a Forwarder Information Base (FIB) that tracks the location (i.e. point of attachment network address) of the mobile terminal. At the forwarder, app updates are addressed using the FIB and forwarded to the mobile terminal.

This system has two update processes that we will track. First, we will track the age $\age(t)$ of the app update process at the mobile. Second, the mobile terminal  sends  ``location updates'' to the forwarder that get written in the FIB. For this process, we wish to track the age $\hat{\age}(t)$ of location updates in the FIB. 

These two age processes are coupled through the FIB. At the forwarder, 
a writer receives location updates from the mobile and writes them to the FIB
while a reader receives  app updates from the application server and needs to read the FIB in order to address the app updates for forwarding to the mobile terminal. 
In short, {\em location updates are written to the FIB and the app updates are client requests to read the FIB}.  

If the mobile terminal has moved and the FIB holds the wrong address, the misaddressed app updates are assumed to be lost and such packet losses will be reflected in increased age in the app updates at the mobile terminal. Misaddressed packets can arise in RCU if the reader reads the FIB while the write of a fresh  location update is in progress. Misaddressed packets occur in RWL when a read lock prevents the writer from writing a fresh location update.  

The app updates in a router's queue are read/service requests to readers. The key idea here is that when the read returns with an address, the queue entity should keep the most recent/freshest read request and forwards that update using the value of returned read. This results in the mobile client receiving the freshest app update. To highlight, the freshness of app updates received at mobile client depends upon two factors: first, if the update was addressed correctly, and second, only the freshest 
app updates are served at the router. Notice that both these factors are inherently affected by the synchronization mechanism being used.  

\subsection{Model Assumptions}
There is no apparent consensus in the literature on modeling random variables associated with the time needed for the execution of a read or write operation. This is indeed the time needed for a software function call that depends upon the underlying hardware and operating system used. In this work, we will assume exponential distributions for both write and read times. If the time of an  operation (read or write) has an exponential $(\mu)$ distribution, then the average time of the operation is $1/\mu$ and we refer to $\mu$ as the speed of the operation.  While exponential models  are decidedly too simple, they enable (with a manageable number of parameters) an analytic characterization  of update age performance induced by the complex RCU and RWL synchronization primitives.

We assume the source submits fresh location updates as rate $\lamhat$ Poisson process to the network and 
that these updates arrive fresh at the FIB writer, i.e. with age $0$. 
The write operations to unlocked memory locations have independent exponential $(\mh)$ service times.
We further assume a preemption in service model; if the writer is busy writing a \locup and a new \locup arrives then the update in service is preempted and writer starts serving this fresher update. Specifically, the writer discards the \locup in service and starts writing the fresh update.

We similarly model the arrivals of client requests (app updates) to the FIB reader as a rate $\lam$ Poisson process, and assume each read request's service/read time is an independent exponential $(\mu)$ random variable.
At the FIB reader, we allow the client requests to be preempted while waiting for the read to return. Specifically, after the read returns, only the fresher \apup is addressed with the FIB read and any previous old update waiting for the read value is preempted. 

We note that modeling and simulation settings are necessarily simplified to facilitate getting some insight into understanding the system. It is possible to 
extend our approach with
more detailed models where a writer takes multiple stages to finish a write, where each stage takes an exponential time, then the total write time PDF is the convolution of these exponential PDFs. However, without the simplified exponential models, the age analysis is intractable, and the alternative is to simulate.
%

\subsection{Paper Outline and Contributions}
\label{sec:paperoutline}

In section~\ref{sec:overview}, we provide an overview of the RCU and RWL synchronization primitives.
In Section~\ref{sec:SHSAoI}, we introduce the
Stochastic Hybrid Systems (SHS) method for AoI evaluation. We
use SHS to compare RCU and RWL in terms of the average age of location and app updates in the update forwarding system introduced in Section~\ref{sec:networkforwarding}.
We show how the app update age process and location update age process are coupled through the FIB. In section \ref{sec:num-eval} we perform numerical evaluations to understand and compare RCU and RWL 
and their effect on location and app update age process. This includes a comparison of  preemptive and non-preemptive RCU and RWL models. While one may speculate that the lock-less RCU approach that enables writing to the FIB without delay should outperform the delay-inducing locks of RWL, our results show that RCU is superior in some operating regimes but worse in others. While these conclusions are specific to our update forwarding example, the approach we develop here can guide the construction of similar comparisons for other distributed updating applications employing shared memory.


\section{Overview of Synchronization Primitives}
\label{sec:overview}
\subsection{Read-Copy-Update (RCU)}
While conventional locking techniques such as Readers-Writer Locks (RWL) enforce strict mutual exclusion between readers and writers in order to prevent destructive modifications, they fall short on concurrent computations by virtue of mutual exclusion. Replacing expensive conventional locking techniques, Read-Copy-Update (RCU) is a synchronization primitive that allows concurrent forward progress for both writers and readers \cite{MckenneyRCU2001}. RCU can be broadly described in two steps \cite{kernelRCU}: 
\begin{enumerate*}
     \item To publish a newer version of a data item\footnote{A writer has \textit{published} an update when it exits its write critical section}, the writer creates a copy of the RCU protected data item, modifies this copy with the newer version of this data item, and atomically replaces the old reference with a reference to this newer version.
     This publishing process runs concurrently with ongoing read processes that continue to read the old copy/version using the old reference. However, new read requests 
     read the most recent 
     version. 
    \item Since some readers in progress hold reference to ``stale'' data, the system defers memory reclamation of old data until after each reader in progress has finished executing its read-side critical section. 
\end{enumerate*}
Therefore, at any point of time, RCU can maintain multiple time-stamped versions of a data item that are concurrently read by readers in the system. 

Every reader that enters a read-side critical section prior to any modification from the writer is able to finish executing its respective critical section. A ``grace period'' starts at the moment the writer publishes the modified data item and enables all RCU read-side critical sections in existence at the beginning of a given grace period to complete~\cite{Desnoyers}. Thus, the end of grace period ensures that it is safe to reclaim the memory and delete the stale copy.


\subsection{Readers-Writer Lock (RWL)}
RWL is a synchronization primitive that enforces mutual exclusion between readers and writers; multiple readers are allowed to read the shared data structure concurrently, while a writer requires exclusive access or a ``lock''\footnote{In shared-memory multiprocessor architectures, a lock is a mechanism that restricts the access to a shared data structure among multiple processors} to that data structure. The focus of most RWL implementations is that no thread should be allowed to starve. Therefore, with just one writer, RWL implementations are mostly \textit{write preferring} \cite{courtois1971}. In this writer priority RWL, once the writer starts waiting in a queue to acquire the lock, the RWL mechanism prevents new readers from acquiring the lock. The writer's acquisition of the lock occurs once all readers already holding the read lock have finished reading. During the write lock, new read lock requests 
are
queued until the writer has released its lock. 
\subsection{RCU and RWL Background}
RCU has been used in a  multitude of places in both user-space and the Linux kernel. For example, in the networking protocol stack, LC Tries employs locking via RCU to enable efficient IP address lookups \cite{NilssonIPAddr}, \cite{lctrie}. User-space RCU  \cite{userspacercu} is used in high-performance DNS servers \cite{knotdns}, in the Linux networking toolkit \cite{netsniff}, in distributed object storage systems \cite{sheepdog}.
Most recently, RCU protected data structures have been employed to ensure wait-free access to machine learning models by inference threads \cite{TensorFlowRCU}. One drawback of a classical RCU mechanism is the \textit{wait-for-readers} (using \textit{synchronize\_rcu()}) primitive where updaters wait for all pre-existing readers to complete their read-side critical sections. Various RCU variants have been proposed (Predicate RCU \cite{predicateRCU}, \cite{gelado2019}, read-log update \cite{matveevRlu}) that address the wait-for-readers problem. Apart from this, \cite{guniguntala} introduced a real-time variant of RCU that allows preemption of read-side critical sections. 

A caveat of RCU is that it doesn't support multiple concurrent updates. A body of research focuses and design of algorithms that support concurrent updates and multi-versioning \cite{ArbelConcurrent2014, matveevRlu, AjitMvrlu}. Further, RCU implementation and verification is non-trivial and several attempts have been made to systematically check the RCU design and code \cite{kokologiannakisstateless,Kokologiannakis2017, liang2018verification, tassarotti2015verifying}. 

Readers-Writer locks are ubiquitous in today's system and are found to support concurrency in virtual file systems, large key-value stores, database systems, software transactional memory implementations \cite{TLRW}. Conventional implementation of Readers-Writer lock suffers from reader-reader scalability and different designs have been proposed for scalable Readers-Writer locks~\cite{dice2019bravo, passiverwl}. Authors in \cite{nir2013numarwl} present the design of a family of RW locks to leverage NUMA features and deliver better performance.

\section{AoI Evaluation of app updates using SHS}

\label{sec:SHSAoI}
\subsection{Stochastic Hybrid Systems (SHS) Overview}
To evaluate AoI of app updates, we use a Stochastic Hybrid Systems (SHS)~\cite{hespanha2006modelling} approach,  a technique introduced for AoI evaluation in \cite{yates2018ToIT} and since employed in AoI evaluation of a variety of status updating systems \cite{Yates-aoi2018,Farazi-KB-aoi2018,Maatouk-AE-aoi2019,Kaul-Yates-isit2018priority,Maatouk-AE-ToN2020,Yates-IT2020,Moltafet-LC-CommLetters2021,Moltafet-LC-ISWCS2021}. 
A stochastic hybrid system has a state-space with two components -- a  discrete component $q(t) \in \Qcal = \{0, 2, \ldots, M\}$ that is a continuous-time finite state Markov Chain and a continuous component $\xv(t) = [x_0(t), \ldots , x_n(t)] \in \R^{n+1}$. In AoI analyses using SHS, each $x_{j}(t) \in \xv(t)$ describes an age process of interest. Each transition $l \in \Lcal$ is a directed edge $(q_l, q'_l)$ with a transition rate $\laml$ in the Markov chain. The age process vector evolves at a unit rate in each discrete state $q \in \Qcal$, i.e., $\frac{d\xv}{dt} = \dot{\xv}(t) = \onev$. A transition $l$ causes a system to jump from discrete state $q_l$ to $q_l^\prime$ and resets the continuous state from $\xv$ to $\xv'$ using a linear transition reset map $\Amat_l \in \{0,1\}^{(n \times n)}$ such that $\xv' = \xv \Amat_l$. For simple queues, examples of transition reset mappings $\set{\Amat_l}$  can be found in \cite{yates2018ToIT}.

For a discrete state $\qbar \in \Qcal$, let $\Lcal_{\qbar}$ and $\Lcal'_{\qbar}$ be sets of incoming and outgoing transitions, i.e.
\begin{align}
    \Lcal_{\qbar} &= \{l \in \Lcal : q'_l = \qbar \}, &
    \Lcal'_{\qbar} &= \{l \in \Lcal : q_l = \qbar \}.
\end{align}
Age analysis using SHS is based on the  expected value processes $\set{\vv_q(t)\colon q\in \Qcal}$ such that 
\begin{align}
 \vv_q(t)
 &=\E{\xv(t)\delta_{q,q(t)}}\nn  &=\rvec{\E{x_1(t)\delta_{q,q(t)}} & \cdots&\E{x_n(t)\delta_{q,q(t)}}},
\end{align}
 with $\delta_{i,j}$ denoting the Kronecker delta function. For the SHS models of age processes considered here, each $\vv_q(t)$ will converge to a fixed point $\vvbar_q$. The fixed points $\set{\vvbar_q\colon q\in\Qcal}$ are the solution to a set of age balance equations. Specifically,
the following theorem provides a simple way to calculate the age balance fixed point and then the average age in an ergodic queueing system. 
\begin{theorem}\thmlabel{AOI-SHS}
\cite[Theorem~4]{yates2018ToIT}
If the discrete-state Markov chain $q(t)\in\Qcal=\set{0,\ldots,M}$ is ergodic 
with stationary distribution 
$\bar{\piv}=\rvec{\bar{\pi}_0 &\cdots &\bar{\pi}_M}>0$ and there exists a 
non-negative vector $\vvbar=\rvec{\vvbar_0&\cdots&\vvbar_M}$
such that 
\begin{align}
\bar{\vv}_{\qbar}\sum_{l\in\Lcal_{\qbar}}\laml &=\onev[]\bar{\pi}_{\qbar}+ \sum_{l\in\Lcal'_{\qbar}}\laml \bar{\vv}_{\ql}\Amat_l,\quad \qbar\in\Qcal,\eqnlabel{AOI-SHS-v}
\end{align}
then
the average age vector is 
$\E{\xv}=\limty{t}\E{\xv(t)}=
\sum_{\qbar\in\Qcal} \vvbar_{\qbar}$.
\end{theorem}

\begin{figure}
\centering
\begin{tabular}{cc}
\begin{minipage}{0.23\textwidth}
\begin{tikzpicture}[->, >=stealth', auto, semithick, node distance=2cm]
\tikzstyle{every state}=[fill=none,draw=black,thick,text=black,scale=0.8]
\node[state]    (4)                     {$4$};
\node[state] (0)[above left of=4] {$0$};
\node[state] (1)[above right of=4] {$1$};
\node[state] (2)[below right of=4] {$2$};
\node[state] (3)[below left of=4] {$3$};
\path
(0) edge[bend left=25, above] node{\small $\lamhat$}  (1)
(1) edge[bend left=25, above] node{\small $\mh$} (0)
(1) edge[loop right] node{\small $\lamhat$} (1)
(1) edge[bend left=25, pos = 0.5, right] node{\small $\lam$} (2)
(2) edge[loop right] node{\small $\lamhat$} (2)
(2) edge[loop below] node{\small $\lam$} (2)
(4) edge[loop below] node{\small $\lam$} (4)
(4) edge[pos = 0.2, above] node{\small $\mu$} (0)
(3) edge[bend right=25, left] node{\small $\mu$} (0)
(0) edge[bend right=25, left] node{\small $\lam$} (3)
(3) edge[loop below] node{\small $\lam$} (3)
(3) edge[bend right=25, below] node{\small $\lamhat$} (2)
(2) edge[bend left = 20,right, midway] node{\small $\mu$} (1)
(4) edge[bend left=20,pos=0.6,left] node{\small $\lamhat$} (2)
(2) edge[bend left=25, below] node{\small $\mh$} (4);
\end{tikzpicture}
\end{minipage} 
& 
\begin{minipage}{0.23\textwidth}
\begin{tikzpicture}[->, >=stealth', auto, semithick, node distance=2cm]
\tikzstyle{every state}=[fill=none,draw=black,thick,text=black,scale=0.8]
\node[state]    (4)                     {$4$};
\node[state] (0)[above left of=4] {$0$};
\node[state] (1)[above right of=4] {$1$};
\node[state] (2)[below right of=4] {$2$};
\node[state] (3)[below left of=4] {$3$};
\path
(0) edge[bend left=25, above] node{\small $\lamhat$}  (1)
(1) edge[bend left=25, above] node{\small $\mh$} (0)
(1) edge[loop right] node{\small $\lamhat$} (1)
(2) edge[loop right] node{\small $\lamhat$} (2)
(2) edge[loop below] node{\small $\lam$} (2)
(0) edge[bend right=25, left] node{\small $\lam$} (3)
(3) edge[bend right=25, left] node{\small $\mu$} (0)
(3) edge[pos = 0.2,right] node{\small $\lamhat$} (4)
(3) edge[loop below] node{\small $\lam$} (3)
(4) edge[pos=0.2, above] node{\small $\mu$} (1)
(1) edge[bend left=25, pos = 0.5, right] node{\small $\lam$} (2)
(4) edge[loop below] node{\small $\lam$} (4)
(4) edge[loop right] node{\small $\lamhat$} (4)
(2) edge[bend left=25, below] node{\small $\mh$} (3);
\end{tikzpicture}
\end{minipage}
\\
{\bf (a)} RCU & {\bf (b)} RWL
\end{tabular}
\caption{SHS Markov chain for {\bf (a)} RCU mechanism and for { \bf (b)} RWL mechanism.}
\label{fig:mc-rcu-rwl}
\vspace{-5mm}
\end{figure}
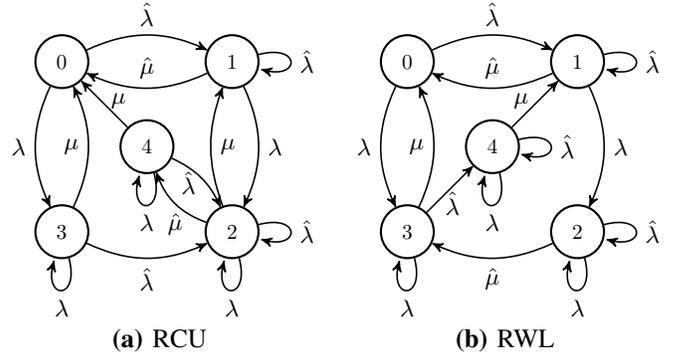

\subsection{RCU and RWL: SHS framework}
We now describe the SHS framework for RCU and RWL considering the packet forwarding example. 
We consider two age vectors: \begin{enumerate*}
    \item the age state of the location update process is $\hat{\xv}(t) = \rvec{\xh{0}(t)&\xh{1}(t)}$, where $\xh{0}(t)$ is the age of the location update seen by the writer and $\xh{1}(t)$ is the age of the current location update in memory;
    \item the age state of the application update process that initiates client read requests is $\xv(t) = \rvec{x_0(t)&x_1(t)}$, where $x_0(t)$ and $x_1(t)$ are  the ages of the most recent application updates at the reader and at the mobile terminal (i.e. the destination monitor) respectively. 
\end{enumerate*}

Since RWL and RCU are fundamentally two different mechanisms for accessing shared memory, the discrete states $\Qcal = \set{0,1,2, 3, 4}$ are similar albeit different: 
\begin{description}
    \item[State 0] The idle state
    \item[State 1] The writer is writing fresh update (with a write lock in RWL)     
    \item[State 2] The writer is writing a fresh update (with a write lock in RWL) but the {\em reader} action is different for RCU and RWL. For RCU, the reader is reading a stale address, but in RWL, the reader has requested a read lock and is waiting for the lock to become active.
    \item[State 3] The reader reading fresh/correct update from memory (with a read lock in RWL)   
    \item[State 4] The reader is reading a stale update (with read lock active in RWL) but the {\em writer} state is different. For RCU, the writer has finished writing the update and this new update is published. For RWL, the writer has requested a write lock and is waiting for the in-progress read to finish.
    \end{description}

\begin{table*}[t]
\centering
$\setlength{\extrarowheight}{0.5mm}
\begin{array}{cc}
\begin{array}[t]{ccccccc}
l & q_l\to q'_l & \laml &  \hat{\xv}\hat{\Amat}_l & \hat{\Amat}_l & \xv\Amat_l & \Amat_l \\ \hline
1 & 0\to 1 & \lamhat & \rvec{0&\xh{1}}&\smvec{0&0\\0&1}&\rvec{x_0&x_1}&\smvec{1&0\\0&1}\\
2 & 1\to 1 & \lamhat & \rvec{0&\xh{1}} & \smvec{0&0\\0&1} & \rvec{x_0&x_1} & \smvec{1&0\\0&1}\\
3 & 2\to 2 & \lamhat & \rvec{0&\xh{1}} & \smvec{0&0\\0&1} & \rvec{x_0&x_1} & \smvec{1&0\\0&1}\\
4 & 3 \to 2 & \lamhat & \rvec{0&\xh{1}} & \smvec{0&0\\0&1}&\rvec{x_0&x_1}&\smvec{1&0\\0&1}\\
5 & 4 \to 2 & \lamhat & \rvec{0&\xh{1}} & \smvec{0&0\\0&1} &\rvec{x_0&x_1}&\smvec{1&0\\0&1}\\
6 & 1\to 0 & \mh & \rvec{\xh{0}&\xh{0}}&\smvec{1&1\\0&0}&\rvec{x_0&x_1}&\smvec{1&0\\0&1}\\
7 & 2 \to 4 & \mh & \rvec{\xh{0}&\xh{0}} & \smvec{1&1\\0&0}&\rvec{x_0&x_1}&\smvec{1&0\\0&1}\\
8 & 0 \to 3 & \lam & \rvec{\xh{0}&\xh{1}} & \smvec{1&0\\0&1}&\rvec{0&x_1}&\smvec{0&0\\0&1}\\
9 & 1\to 2 & \lam & \rvec{\xh{0}&\xh{1}}&\smvec{1&0\\0&1}&\rvec{0&x_1}&\smvec{0&0\\0&1}\\
10 & 2\to 2 & \lam & \rvec{\xh{0}&\xh{1}} & \smvec{1&0\\0&1} & \rvec{0&x_1} & \smvec{0&0\\0&1}\\
11 & 3\to 3 & \lam & \rvec{\xh{0}&\xh{1}} & \smvec{1&0\\0&1} & \rvec{0&x_1} & \smvec{0&0\\0&1}\\
12 & 4\to 4 & \lam & \rvec{\xh{0}&\xh{1}} & \smvec{1&0\\0&1} & \rvec{0&x_1} & \smvec{0&0\\0&1}\\
13 & 3 \to 0 & \mu & \rvec{\xh{0}&\xh{1}} & \smvec{1&0\\0&1}&\rvec{x_0&x_0}&\smvec{1&1\\0&0}\\
14 & 4 \to 0 & \mu & \rvec{\xh{0}&\xh{1}} & \smvec{1&0\\0&1}&\rvec{x_0&x_1}&\smvec{1&0\\0&1}\\
15 & 2 \to 1 & \mu & \rvec{\xh{0}&\xh{1}} & \smvec{1&0\\0&1}&\rvec{x_0&x_1}&\smvec{1&0\\0&1}
\end{array}& 
\setlength{\extrarowheight}{0.5mm}
\begin{array}[t]{ccccccc}
l & q_l\to q'_l & \laml &  \hat{\xv}\hat{\Amat}_l & \hat{\Amat}_l & \xv\Amat_l & \Amat_l \\ \hline
1 & 0\to 1 &\lamhat	& \rvec{0&\xh{1}}&\smvec{0&0\\0&1}&\rvec{x_0&x_1}&\smvec{1&0\\0&1}\\
2 & 1\to 1 & \lamhat &\rvec{0&\xh{1}}&\smvec{0&0\\0&1}&\rvec{x_0&x_1}&\smvec{1&0\\0&1}\\
3 & 2\to 2 & \lamhat &
\rvec{0&\xh{1}}&\smvec{0&0\\0&1}&\rvec{x_0&x_1}&\smvec{1&0\\0&1}\\
4 & 3\to 4 & \lamhat & 
\rvec{0&\xh{1}}&\smvec{0&0\\0&1}&\rvec{x_0&x_1}&\smvec{1&0\\0&1}\\
5 & 4\to 4 & \lamhat & 
\rvec{0&\xh{1}}&\smvec{0&0\\0&1}&\rvec{x_0&x_1}&\smvec{1&0\\0&1}\\
6 & 1\to 0 & \mh & \rvec{\xh{0}&\xh{0}}&\smvec{1&1\\0&0}&\rvec{x_0&x_1}&\smvec{1&0\\0&1}\\
7 & 2\to 3 & \mh &
\rvec{\xh{0}&\xh{0}}&\smvec{1&1\\0&0}&\rvec{x_0&x_1}&\smvec{1&0\\0&1}\\
8 & 0\to 3 & \lam &
\rvec{\xh{0}&\xh{1}}&\smvec{1&0\\0&1}&\rvec{0&x_1}&\smvec{0&0\\0&1}\\
9 & 1\to 2 & \lam & 
\rvec{\xh{0}&\xh{1}} & \smvec{1&0\\0&1}&\rvec{0&x_1}&\smvec{0&0\\0&1}\\
10 & 2\to 2 & \lam &
\rvec{\xh{0}&\xh{1}}&\smvec{1&0\\0&1}&\rvec{0&x_1}&\smvec{0&0\\0&1}\\
11 & 3\to 3 & \lam &
\rvec{\xh{0}&\xh{1}}&\smvec{1&0\\0&1}&\rvec{0&x_1}&\smvec{0&0\\0&1}\\
12 & 4\to 4 & \lam &
\rvec{\xh{0}&\xh{1}}&\smvec{1&0\\0&1}&\rvec{0&x_1}&\smvec{0&0\\0&1}\\
13 & 3\to 0 & \mu &
\rvec{\xh{0}&\xh{1}}&\smvec{1&0\\0&1}&\rvec{x_0&x_0}&\smvec{1&1\\0&0}\\
14 & 4\to 1 & \mu &
\rvec{\xh{0}&\xh{1}}&\smvec{1&0\\0&1}&\rvec{x_0&x_1}&\smvec{1&0\\0&1}\\
\end{array}\vspace{2mm}\\
\text{{\bf (a)} RCU} & \text{{\bf (b)} RWL}
\end{array}$
\caption{SHS transitions 
for tracking age in Markov chains of Fig.~\ref{fig:mc-rcu-rwl} for (a) RCU and (b) RWL.}
\label{tab:shs-transition}
\end{table*}

The discrete-state Markov chains for RCU and RWL are shown in Fig.~\ref{fig:mc-rcu-rwl}(a) and~\ref{fig:mc-rcu-rwl}(b) respectively.
The SHS transition reset maps for RCU and RWL are shown in 
Tables~\ref{tab:shs-transition}(a) and \ref{tab:shs-transition}(b).
In each table, a transition $l$ from state $q_l$ to $q'_l$ occurs at rate $\laml$ with age reset maps \begin{align}\eqnlabel{Amat-pair}
\xv' &= \xv \Amat_l, 
& 
\hat{\xv}'=\hat{\xv}\hat{\Amat}_l.
\end{align}
\Eqnref{Amat-pair} highlights how the the app update process $\xv(t)$ and location update process $\hat{\xv}(t)$ are coupled only through the changes in the Markov chain at the forwarder. 
 In each transition $l$,  either the app update process $\xv(t)$ changes or the location update process $\hat{\xv}(t)$ changes, but not both. That is, either $\Amat_l$ or $\hat{\Amat}_l$ is an identity matrix for each transition $l$. 


\subsection{RCU and RWL: SHS Transitions}
\label{sec:rcu-rwl-shs}
Here we describe the SHS transitions for both RCU and RWL that are enumerated in Tables~\ref{tab:shs-transition}(a) and~\ref{tab:shs-transition}(b). 
  For each collection of transitions, we focus on the age state process ($\xv(t)$ or $\hat{\xv}(t)$) that changes. In particular, we first describe what is common to both RCU and RWL. This is followed by details specific to RCU and RWL respectively.  

\begin{itemize}
\item  {$l=1,\ldots,5$:} In each state $0,\ldots,4$, the writer receives a fresh location update and initiates the write mechanism. Since the location update is fresh, $\xh{0}'=0$ whereas $\xh{1}'=\xh{1}$ is unchanged as the location update  has not yet been written to the FIB. In transitions $l=2,3$, the writer preempts an in-progress write with an updated location.
\begin{description}
    \item[RCU] Following transition $l=5$,   the in-progress read will now be returning an outdated address. 
\item[RWL] In transition $l=1$, the writer acquires a write-lock. In transitions $l=2,3$, the writer already holds the write-lock.
In transitions $l=4,5$, the writer requests a write lock but the request is queued as the reader is in a critical section.
\end{description}
\end{itemize}
\begin{itemize}
    \item $l = 6,7$: 
     The writer finishes writing to the FIB and publishes a new location update;  $\xh{0}' = \xh{0}$ is unchanged since no new location update arrives at the writer but $\xh{1}' = \xh{0}$ as the age at the FIB is reset to the age of the just-written update.  In transition $l=6$, the system goes to the idle state.
\begin{description}
\item[RCU] In transition $l=7$, the system goes to state $4$ because a grace period starts with a read in progress.
    \item[RWL]For transition, $l=7$, there is a pending read request and so the reader acquires the read lock and enters a read-side critical section.
\end{description}     
     
\item $l = 8,\ldots,12$:
    In each discrete state $0,\ldots,4$, an app update arrives, initiating a read request; $x'_0 = 0$ as the app update is fresh at the reader but $x'_1 = x_1$ since the app update has not yet been delivered to the mobile terminal. 
    
\begin{description}   
    \item[RCU] In transitions $l=9,10$, the system enters state $2$ in which the writer is simultaneously writing a fresh location update. {\em Consequently, the reader will read a stale address from the FIB in state $2$.} For transitions $l=11,12$, a read was already in-progress; when that read completes, the address returned by the FIB is used to address this most recent app update. Effectively, the arriving app update preempts the prior update that had been held by the reader. 
    
\item[RWL] In transition $l=8$, the reader immediately acquires a read-lock on the FIB and initiates the read. In transition $l=9$, the app update arrives in a write-lock state, the reader requests a read lock on the FIB that is denied and the system transitions to state $2$, the write-lock with read pending state. In transition $l=10$, the fresh app update arrives in state $2$ and the system stays in the write-lock with read-pending state. In transitions $l=11,12$, the fresh app update arrives in a state with the read-lock already active. Hence, in transitions $l=10,11,12$, the system remains in its same state but the fresh app update preempts the prior app update at the reader that was waiting to be addressed and sent. We note that following transition $l=10$ or $l=11$, there is the chance that the app update will eventually be correctly delivered to the mobile. However, in the case of transition $l=12$, the app update, if not preempted,  will eventually read an outdated address and go misaddressed.
  \end{description}
    
    \item $l = 13$:
    The reader retrieves a location address from the FIB and exits the critical section; $x'_0 = x_0$ but $x'_1 = x_0$ as the age at mobile terminal is reset to the age of the app update that was just addressed and delivered to the mobile. In both RCU and RWL, the system returns to the idle state.
    
    \item $l = 14$:
    The reader retrieves a stale location address from the FIB, exits the critical section and attempts to forward the app update to the mobile; 
    $x'_0 = x_0$ but  $x'_1 = x_1$ is unchanged, i.e. the age at mobile is not reset since the address read is outdated and the misaddressed app update is lost in transit.
\begin{description}
    \item[RCU] When this transition occurs, the writer is idle, having already finished writing its location update to the FIB. However, the  reader is fetching the prior copy holding the outdated location.

    \item[RWL] When this transition occurs, the writer has received the location update, but is in a write-pending state waiting for the read-lock to be released.
\end{description}
    
    \item $l=15$: 
    When this transition occurs, an RCU read finishes while an in-progress RCU write is updating the FIB with a location update that occurred while the read was in progress.  (This transition is exclusive to RCU since RWL locking prohibits simultaneous reading and writing.)   Similar to transition $l=14$, the reader retrieves the stale (prior) location address and attempts to forward the app update to the mobile.  Since this address is outdated, the misaddressed app update is lost in transit;  $x'_0 = x_0$ and  $x'_1 = x_1$ are unchanged.
\end{itemize}
\subsection{RCU: SHS Age Analysis}
\label{sec:rcu-analysis}
For the SHS analysis, we  employ  the normalized rates \begin{equation}
    \rh={\lamhat}/\mh,\quad \bt={\lam}/{\mh},\quad \sig=\mu/\mh.
\end{equation} 
We note that $\rh$ is the offered load of location updates being written to the FIB. Similarly, $\bt$ is the normalized arrival rate of FIB read requests.
For RCU, the  Fig.~\ref{fig:mc-rcu-rwl}(a) Markov chain has stationary probabilities $\piv =\rvec{\pi_0 & \pi_1 & \pi_2 & \pi_3 & \pi_4}$ with normalization constant $C_\pi$ given by
\begin{subequations}
\eqnlabel{rcustatprob}
\begin{align}
    \piv &= C_\pi^{-1} \rvec{\sig & \rh \sig & \bt \rh &  \frac{\bt \sig}{\rh+\sig} & \frac{\bt\rh}{\rh+\sig}},\\
    C_\pi &= (1+\rh)(\bt+\sig).
\end{align}
\end{subequations}

We now apply \Thmref{AOI-SHS} to  the SHS reset maps 
in Table~\ref{tab:shs-transition}(a). With the shorthand notations 
\begin{equation}
    \lamsum=\lambda+\lamhat,\qquad \musum =\mu+\mh,
    \end{equation}
From \Thmref{AOI-SHS}, the RCU location update age process $\hat{\xv}(t)$ has age balance fixed points $\vvhat_q=\rvec{\vhat_{q0}&\vhat_{q1}}$ satisfying
\begin{subequations}
\eqnlabel{rculocationshs}
\begin{align}
    \lamsum\vvmhat{0} &= \onev[]\pibar_0 + \mh \vvmhat{1}\hat{\Amat}_6 + \mu \vvmhat{3}\hat{\Amat}_{13} + \mu\vvmhat{4}\hat{\Amat}_{14}, \\
    (\lamsum+\mh)\vvmhat{1} &= \onev[] \pibar_1 + \lamhat \vvmhat{0} \hat{\Amat}_1 + \lamhat \vvmhat{1}\hat{\Amat}_{2} + \mu\vvmhat{2}\hat{\Amat}_{15}, \\
    (\lamsum+\musum)\vvmhat{2} &= \onev[]\pibar_2 +  \lamhat\vvmhat{2}\hat{\Amat}_{3} + \lamhat\vvmhat{3}\hat{\Amat}_{4}
    +\lamhat\vvmhat{4}\hat{\Amat}_{5} \nn
    &\qquad+
    \lam\vvmhat{1}\hat{\Amat}_{9} +
    \lam\vvmhat{2}\hat{\Amat}_{10}, \\
    (\lamsum +\mu)\vvmhat{3} &= \onev[]\pibar_{3} + \lam \vvmhat{0}\hat{\Amat}_8 + \lam \vvmhat{3}\hat{\Amat}_{11}, \\
    (\lamsum+\mu)\vvmhat{4} &= \onev[]\pibar_{4} + \mh\vvmhat{2}\hat{\Amat}_7 + \lam\vvmhat{4}\hat{\Amat}_{12}. 
\end{align}
\end{subequations}
From Table~\ref{tab:shs-transition}(a) we see that  $\hat{\Amat}_8,\hat{\Amat}_9,\ldots,\hat{\Amat}_{15}$ are all identity matrices. It follows that \eqnref{rculocationshs} simplifies to
\begin{subequations}
\eqnlabel{rculocationshs-noident}
\begin{IEEEeqnarray}{rCl}
    \lamsum\vvmhat{0} &=& \onev[]\pibar_0 + \mh \vvmhat{1}\hat{\Amat}_6 + \mu \vvmhat{3} + \mu\vvmhat{4}, \\
    (\lamsum+\mh)\vvmhat{1} &=& \onev[] \pibar_1 + \lamhat \vvmhat{0} \hat{\Amat}_1 + \lamhat \vvmhat{1}\hat{\Amat}_{2} + \mu\vvmhat{2}, \\
    (\lamhat+\musum)\vvmhat{2} &=& \onev[]\pibar_2 +  \lamhat\vvmhat{2}\hat{\Amat}_{3} + \lamhat\vvmhat{3}\hat{\Amat}_{4}
    +\lamhat\vvmhat{4}\hat{\Amat}_{5} 
    + \lam\vvmhat{1},\IEEEeqnarraynumspace \\
    (\lamsum +\mu)\vvmhat{3} &=& \onev[]\pibar_{3} + \lam \vvmhat{0} + \lam \vvmhat{3}, \\
    (\lamsum+\mu)\vvmhat{4} &=& \onev[]\pibar_{4} + \mh\vvmhat{2}\hat{\Amat}_7 + \lam\vvmhat{4}. 
\end{IEEEeqnarray}%
\end{subequations}

Because the RCU app updates employ the same discrete state Markov chain as the RCU location updates, the RCU age balance equations for the app updates are identical to \eqnref{rculocationshs} with 
$\vvmhat{0},\ldots,\vvmhat{4}$ and $\hat{\Amat}_1,\ldots,\hat{\Amat}_{15}$ replaced by 
$\vvbar_{0},\ldots,\vvbar_{4}$ and $\Amat_1,\ldots,\Amat_{15}$ respectively:
\begin{subequations}\eqnlabel{rcuappshs}
\begin{align}
    \lamsum\vvbar_{0} &= \onev[]\pibar_0 + \mh \vvbar_{1}\Amat_6 + \mu \vvbar_{3}\Amat_{13} + \mu\vvbar_{4}\Amat_{14}, \\
    (\lamsum+\mh)\vvbar_{1} &= \onev[] \pibar_1 + \lamhat \vvbar_{0} \Amat_1 + \lamhat \vvbar_{1}\Amat_{2} + \mu\vvbar_{2}\Amat_{15}, \\
    (\lamsum+\musum)\vvbar_{2} &= \onev[]\pibar_2 +  \lamhat\vvbar_{2}\Amat_{3} + \lamhat\vvbar_{3}\Amat_{4}
    +\lamhat\vvbar_{4}\Amat_{5} \nn
    &\qquad+
    \lam\vvbar_{1}\Amat_{9} +
    \lam\vvbar_{2}\Amat_{10},\\
    (\lamsum +\mu)\vvbar_{3} &= \onev[]\pibar_{3} + \lam \vvbar_{0}\Amat_8 + \lam \vvbar_{3}\Amat_{11}, \\
    (\lamsum+\mu)\vvbar_{4} &= \onev[]\pibar_{4} + \mh\vvmbar{2}\Amat_7 + \lam\vvmbar{4}\Amat_{12}. 
\end{align}
\end{subequations}
For these equations, we note from Table~\ref{tab:shs-transition}(b) that $\Amat_1,\ldots,\Amat_7$ and $\Amat_{14},\Amat_{15}$ are all identity matrices and this will lead to a set of simplified age balance equations, equivalent to \eqnref{rculocationshs-noident} for the RCU location updates.
Numerical evaluation of $\vvmhat{0},\ldots,\vvmhat{4}$ and 
$\vvbar_{0},\ldots,\vvbar_{4}$ using \eqnref{rculocationshs-noident} and the simplified version of \eqnref{rcuappshs} respectively is straightforward. 
It follows from \Thmref{AOI-SHS}  that average age $\E{\hat{\age}}$  of a location update  in the FIB and the average age $\E{\age}$ of an app update at the mobile terminal are
\begin{align}
\E{\hat{\age}} &=
\sum_{q=0}^4\vhat_{q1}, & 
\E{\age} &=
\sum_{q=0}^4\vbar_{q1}.
\eqnlabel{app-average-age}
\end{align}


\subsection{RWL: SHS  Age Analysis}
\label{sec:RWL-analysis}

The RWL Markov chain in Fig.~\ref{fig:mc-rcu-rwl}(b) has stationary probabilities $\piv$
with normalization constant $C_\pi$ given by
\begin{subequations}
\begin{align}
    \piv &=  \rvec{\pi_0 & \pi_1 & \pi_2 & \pi_3 & \pi_4}= C_\pi^{-1} \begin{bmatrix}\sig(\rh+\sig+\bt\sig)\\
    \rh\sig(\bt+\rh+\sig) \\ \bt\rh\sig(\bt+\rh+\sig)\\ \bt\sig(1+\bt+\rh)\\ \bt\rh(1+\bt+\rh)\end{bmatrix}^\top\\
C_\pi &= \bt\rh(1+\bt+\rh) + \sig(1+\bt)(1+\rh)(\sig+\bt+\rh).
\end{align}
\end{subequations}
The age balance equations based on the SHS reset maps shown in Table~\ref{tab:shs-transition}(b) for RWL location updates are:
\begin{subequations}
\eqnlabel{rwllocationshs}
\begin{align}
    \lamsum\vvmhat{0} &= \onev[]\pibar_0 + \mh \vvmhat{1}\hat{\Amat}_6 + \mu \vvmhat{3}\hat{\Amat}_{13}, \\
    (\lamsum+\mh)\vvmhat{1} &= \onev[] \pibar_1 + \lamhat \vvmhat{0} \hat{\Amat}_1 + \lamhat \vvmhat{1} \hat{\Amat}_2 + \mu\vvmhat{4}\hat{\Amat}_{14}, \\
    (\lamsum+\mh)\vvmhat{2} &= \onev[]\pibar_{2} + \lamhat \vvmhat{2} \hat{\Amat}_3 + \lam\vvmhat{1}\hat{\Amat}_9 + \lam \vvmhat{2} \hat{\Amat}_{10}, \\
    (\lamsum+\mu)\vvmhat{3} &= \onev[]\pibar_{3} + \mh \vvmhat{2} \hat{\Amat}_7+\lam \vvmhat{0}\hat{\Amat}_8 + \lam \vvmhat{3}\hat{\Amat}_{11}, \\
    (\lamsum+\mu)\vvmhat{\gp} &= \onev[]\pibar_{4} + \lamhat\vvmbar{3}\hat{\Amat}_4 + \lamhat\vvmbar{4}\hat{\Amat}_5 + \lam\vvmbar{4}\hat{\Amat}_{12}. 
\end{align}
\end{subequations}
For the app updates described by the age process $\xv(t)$, there is a set of SHS equations identical to \eqnref{rwllocationshs},  but with transition reset maps $\Amat_l$ in place of $\hat{\Amat}_l$ and variables $\vv_q = \rvec{v_{q0}&v_{q1}}$ in place of $\vvmhat{q} = \rvec{\vhat_{q0}&\vhat_{q1}}$.
Once again, we solve these equations numerically and it follows from \Thmref{AOI-SHS}  that average age $\E{\age}$ of the  RWL app update process  is given by \eqnref{app-average-age}.

The SHS analysis of RCU and RWL corresponding to Equations~\eqnref{rcustatprob}
through~\eqnref{rwllocationshs} incorporate preemption of updates at the FIB reader and writer. We will also consider non-preemptive  versions of RCU and RWL. In these systems, app updates arriving during the reader's busy state are discarded. Consequently, when a read returns with an address, it addresses the app update that initiated the read request.  

In SHS models of these non-preemptive systems, the states of Markov chains in Fig.~\ref{fig:mc-rcu-rwl} remain unchanged. However, the self transitions of rate $\lam$ are absent. In particular, the SHS transition tables of the non-preemptive RCU and RWL systems are given in Table~\ref{tab:shs-transition} except transitions $l=10,11,12$ in both Table \ref{tab:shs-transition}(a) and Table \ref{tab:shs-transition}(b) are deleted. The age balance equations then can be obtained from \Thmref{AOI-SHS} in the same way that we derived \eqnref{rculocationshs}. We don't explicitly enumerate these age balance equations here; however, in section \ref{sec:num-eval}, we numerically compare the performance of preemptive and non-preemptive systems. 

\section{Numerical Results}
\label{sec:num-eval}
We note that while RCU writes 
are lock-less, they  
can be
still heavy as the writer tracks the start and end of a grace period, and is also responsible for memory reclamation of stale copies.
On the other hand, RWL writes use locks to update the data structure. 
Locking requires expensive atomic operations such as compare-and-swap and thus the corresponding software functionality tends to run slow \cite{hartrcuperf}.  We characterize the RCU and RWL write speeds by the exponential rate parameters $\mh_{\RCU}$ and $\mh_{\RWL}$; however, it is ambiguous whether $\mh_{\RCU} > \mh_{\RWL}$ or vice-versa. Thus, in order to focus on the effects of other system parameters, our numerical evaluations assume \begin{equation}
    \mh_{\RWL} =\mh_{\RCU} =  \mh.
\end{equation} 
In contrast to the ambiguity associated with relative write speeds,  
reads in RCU are typically fast, sometimes an order of magnitude faster than uncontended locking \cite{Desnoyers}. In our SHS models,  read rates are characterized by parameter $\mu$ and since read side primitives are lighter (i.e. faster) in RCU, this corresponds to  
$\mu_{\RCU} > \mu_{\RWL}$. 

With the definition of the normalized rates 
\begin{equation}\eqnlabel{normalization}
\rh=\frac{\lamhat}{\mh},\quad \bt=\frac{\lam}{\mh},\quad
\sig_{\RWL} = \frac{\mu_{\RWL}}{\mh},\quad \sig_{\RCU}=\frac{\mu_{\RCU}}{\mh},
    \end{equation}
we now present some results 
from numerically evaluating \eqnref{rculocationshs-noident}, \eqnref{rcuappshs}, \eqnref{rwllocationshs} with $\mh=1$. 
Hence age will be measured in the units of $1/\mh$, the average shared memory write time. 
As explained, our numerical evaluations consider cases with $\sig_{\RWL}<\sig_{\RCU}$, along with the further assumption $\sig_{\RCU}=10$, corresponding to RCU reads being $10\times$ faster than RCU writes.

In addition, our evaluations will vary $\rh$ over the interval $[0,0.1]$. At $\rh=0$, the mobile terminal is stationary and never changes its network address. On the other hand,  the upper limit $\rh=0.1$  represents an extreme value in that the average time between location changes $1/\lamhat$ is only $10\times$ longer than the average time $1/\mh$ to write to shared memory. For example, very slow memory writes requiring time $1/\mh=1$~ms would correspond to $\lamhat=0.1$ location changes per millisecond, or $100$ location changes per second. While this would be  an extreme level of user mobility in a traditional wireless network environment, there may be other network scenarios in which the mobile user is perhaps a software agent, for which this is appropriate. 
With these constraints,
we aim to provide an 
informative
comparison between RCU and RWL systems. 

\begin{figure}[t]
    \centering
    \begin{tabular}{c}
       \includegraphics{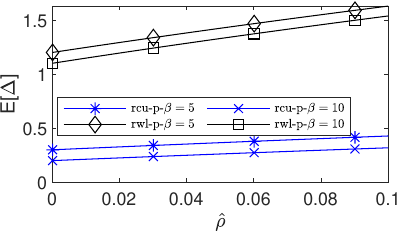} \\
          \textbf{(a)} \\
        \includegraphics{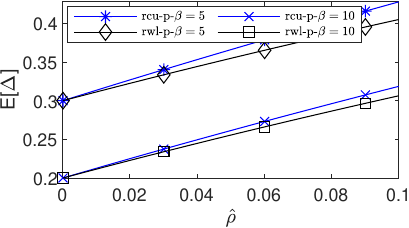} \\
        \textbf{(b)}
    \end{tabular}
    \caption{AoI at mobile client when using RCU preemption (rcu -p) and RWL preemption (rwl-p) as a function of normalized write request rate $\rh=\lamhat/\mh$, against different values of normalized read rate $\beta=\lam/\mh$ and  $\sig_{\RCU} = 10$  with \textbf{(a)} $\sig_{\RWL} = 1$,
    and \textbf{(b)} $\sig_{\RWL}=10$.}
    \label{fig:rcuvsrwl_appage}
\end{figure}

In Fig.~\ref{fig:rcuvsrwl_appage}, we plot the average app age $\E{\age}$ as a function of $\rh$.
A larger $\rh$ means that the mobile is moving faster and changing its location more frequently, and so more app update packets are misaddressed, resulting in increased app age at the mobile terminal. In the same figure, we notice the effect of slower reads in RWL on app age. The app age at the mobile client increases in proportion to the service time of the app updates at the forwarder. Additionally, a slower read with a read lock activated corresponds to the writer being locked out without being able to write a fresher location update.

On the other hand, timely updating is achieved with RCU's fast read-side primitives and shown in Fig.~\ref{fig:rcuvsrwl_appage}(a). We also note that an RWL system with fast reads, say $\sig_{\RWL} = 10$, performs better than RCU with $\sig_{\RCU} = 10$, especially at higher values of $\rh$; see Fig.~\ref{fig:rcuvsrwl_appage}(b). In this case, larger $\rh$ corresponds to a greater likelihood that the FIB address is outdated, but an exclusive write lock prevents the reader from reading a stale address. The lock-less operation of RCU enables the reader to read the outdated FIB. We note that our analysis and numerical evaluations align with RCU literature that RCU is not suitable for update heavy scenarios. 

From the Markov chains in Fig.~\ref{fig:mc-rcu-rwl}, it is also instructive to evaluate the probability an app update is delivered. 
For RCU, an app update arriving in state $0$ or state $3$ is delivered with probability $\mu/(\lamsum+\mu)$, which is the probability that the address read required by the app update finishes before a location update occurs or gets preempted by a fresher app update. 
App updates arriving in states $1$, $2$, and $4$ are discarded primarily because the read request initiated by the updates will return a stale address as the writer is writing a fresher update in each of these states. Thus, the probability  that an app update is delivered under RCU is 
\begin{equation}
    P_{\RCU} = (\pi_0+\pi_3)\frac{\mu}{\lamsum+\mu}.
    \eqnlabel{rcuprobdelivery}
\end{equation}

\begin{figure}[t]
    \centering
    \begin{tabular}{c}
\includegraphics{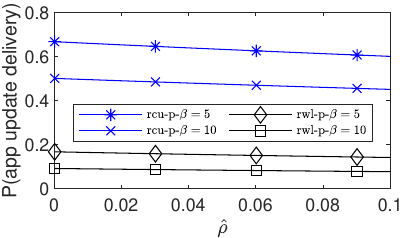} \\
         \textbf{(a)} \\         \includegraphics{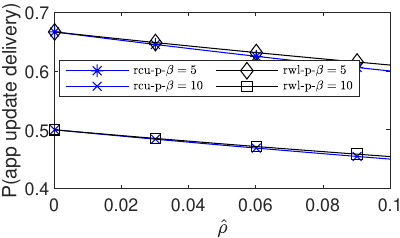} \\
         \textbf{(b)}
    \end{tabular}
    \caption{Probability that an app update arriving at router is delivered correctly when $\sig_{\RCU} = 10$ and when \textbf{(a)} $\sig_{\RWL}=1$ and \textbf{(b)} $\sig_{\RWL}=10$.}
    \label{fig:prob-app-dv}
    \vspace{-5mm}
\end{figure}

For RWL, app updates arriving in state $0$ or state $3$ are delivered with same probability as RCU, i.e. $\mu/(\lamsum+\mu)$. App updates arriving in state $1$ or state $2$ are delivered with probability 
$[{\mh}/{(\mh+\lam)}][{\mu}/{(\lamsum+\mu)}]$.
Thus, the probability that an app update is delivered  is 
\begin{equation}
    P_{\RWL} = (\pi_0+\pi_1 \frac{\mh}{\mh+\lam}+\pi_2\frac{\mh}{\mh+\lam}+\pi_3)\frac{\mu}{\lamsum+\mu}.
    \eqnlabel{rwlprobdelivery}
\end{equation}
Fig.~\ref{fig:prob-app-dv} shows that $P_{\RCU}$ and $P_{\RWL}$ in \eqnref{rcuprobdelivery} and \eqnref{rwlprobdelivery} decrease as a function of normalized write request rate.   
In comparing Figs~\ref{fig:rcuvsrwl_appage} and~\ref{fig:prob-app-dv}, we see that for both RCU and RWL that the average age $\E{\age}$ becomes worse as the delivery probability decreases. 

\begin{figure}[t]
    \centering
    \begin{tabular}{c}  \includegraphics{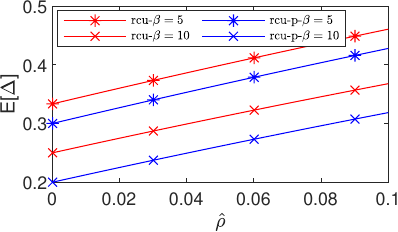} \\
        \textbf{(a)} \\
\includegraphics{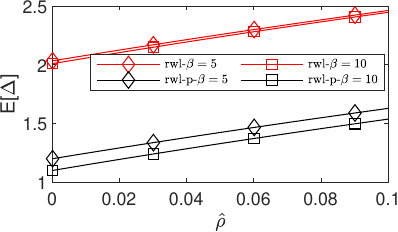} \\
        \textbf{(b)}
    \end{tabular}
    \caption{AoI performance with and without preemption for \textbf{(a)} RCU with $\sig_{\RCU}=10$, and \textbf{(b)} RWL with $\sig_{\RWL} = 1$.}
    \label{fig:pvsnp}
    \vspace{-5mm}
\end{figure}

Fig.~\ref{fig:pvsnp} demonstrates the timeliness gain achieved by employing preemption of app updates held by the reader. For $\sig_{\RCU}=10$, $\sig_{\RWL}=1$, and $\bt=10$, this gain is almost $15\%$ for RCU and $45\%$ for RWL. From the AoI perspective, preemption helps more in RWL as it allows a slower read to service the most recent app update. Nevertheless, we note from Fig.~\ref{fig:pvsnp} 
that preemption mechanisms generally reduce AoI. 

\begin{figure}[t]
    \centering
\includegraphics{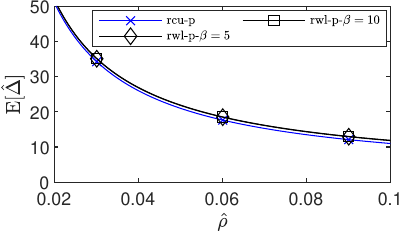}
    \caption{AoI at memory when $\sig_{\RCU}=10$ and $\sig_{\RWL}=1$.}
    \label{fig:rcuvsrwl_locage}
    \vspace{-5mm}
\end{figure}

In Fig.~\ref{fig:rcuvsrwl_locage}, we observe that the age of location updates in the memory is $\E{\hat{\age}}\approx 1/\rh$ for both RCU and RWL. This demonstrates that essentially all location updates are promptly stored in memory and that $\E{\hat{\age}}$ is dominated by the relatively low frequency of location changes. This is the exception to the customary assumption that system performance improves with decreasing age. In this case, 
increasing the rate of location changes reduces the age of location updates in memory, but it also increases the probability that app updates go misaddressed. In this system, the timeliness of location updates would be better described  using metrics such as Age of Incorrect Information \cite{Maatouk-KA-TNet2020} or Age of Synchronization \cite{Cho-GM-tods2003effective,Zhong-YS-isit2018} that account for whether the current update is correct.

\section{Conclusion}\label{sec:conclusion}

This work has explored the impact of synchronization primitives on timely updating.
We modeled and developed a packet forwarding 
scenario in which location updates from a mobile terminal  are written to a forwarding table  and application updates need to read the forwarding table in order to ensure their correct addressing for delivery to a mobile terminal. In this system, we saw the tension between writer and reader, both in the analytic models and in the 
corresponding numerical evaluations.
While timeliness of the location updates in the table is desirable, excessive updating can be at the expense of timely reading of the table. We now discuss some of the many open questions and issues related to this work.

In this work, we considered only a tightly-coupled source and writer in which fresh (zero age) updates are delivered to the writer. However, there are many physical situations in which a loosely coupled source and writer would be appropriate. For example, when the source is a camera sensor and the update is an image, both image processing at the sensor and transmission  of the image  to the writer would contribute to the update preparation time. It is obvious that this additional latency would contribute directly to the age of updates written to memory. What is perhaps less obvious is that this should prompt the writer to be  parsimonious in writing.  In the AoI literature, there is evidence \cite{Sun-UBYKS-IT2017UpdateorWait,Yates-isit2015} that delaying new updates when the current update is relatively fresh can be age-optimal. The insight is that one should not commit system resources to producing a new update when it offers only a small age reduction  relative to the current update. While the setting here is different, it seems likely that similar ideas would also be age-reducing in writing to shared memory.


We have also assumed in this work that a reader does not maintain a local cache copy of its most recent read. With a local cache, the reader can fulfill a read request by either returning the cached update or by requesting a new read lock to return a potentially fresher read from shared memory. While this optimization may improve timeliness of the delivered read, it also highlights a fundamental difference between age and latency. 
In responding to a client with a local copy,  latency, as measured by the {\em turnaround time},  is reduced since the reader has unrestricted access to its local cache. On the other hand, 
a response that  reads the shared memory is likely to be fresher. However, by virtue of mutual-exclusion, the reader may wait to complete its read of the shared memory, thus increasing the turnaround time to the client.  In fact, the reader can optimize its decision making, possibly with age-dependent policies, and this will induce an  \textit{age-latency tradeoff} that needs to be explored.

We observe that RWL and synchronization primitives in general  admit a combinatorial explosion of system models  in specifying the behaviour of readers and writers. We have already described how the source-writer may be loosely or tightly coupled, how the reader may or may not maintain a local cache, and how both reader and writer may or may not employ update preemption mechanisms. Furthermore, conclusions of this work are tightly coupled to our update forwarding scenario.  There is considerable work to be done in the exploring timeliness in other applications and systems employing shared memory. 




\bibliographystyle{IEEEtran}
\bibliography{output-aoi, output-refs}


\end{document}